# Mechanics of bioinspired fiber reinforced elastomers

Aritra Chatterjee[1], Nimesh R. Chahare[2], Paturu Kondaiah[3], Namrata Gundiah[1,2*]

[1]Centre for Biosystems Science and Engineering,

[2]Department of Mechanical Engineering,

[3]Department of Molecular Reproduction, Development and Genetics,

Indian Institute of Science, Bangalore 560012, Karnataka

* For correspondence:

Namrata Gundiah, Department of Mechanical Engineering

Indian Institute of Science, Bangalore 560012.

**Email**: namrata@iisc.ac.in, ngundiah@gmail.com

**Tel (Off/ Lab)**: 91 80 2293 2860/ 3366

**Abstract**

Fiber reinforcement is a crucial attribute of soft bodied muscular hydrostats that have the ability to undergo large deformations and maintain their posture. Helically wound fibers around the cylindrical worm body help control the tube diameter and length. Geometric considerations show that a fiber winding angle of 54.7°, called the *magic angle*, results in a maximum enclosed volume. Few studies have explored the effects of differential fiber winding on the large deformation mechanics of fiber reinforced elastomers (FRE). We fabricated FRE materials in transversely isotropic layouts varying from 0-90° using a custom filament winding technique and characterized the nonlinear stress-strain relationships using uniaxial and equibiaxial experiments. We used these data within a continuum mechanical framework to propose a novel constitutive model for incompressible FRE materials with embedded extensible fibers. The model includes individual contributions from the matrix and fibers in addition to coupled terms in strain invariants, $I_1$ and $I_4$. The deviatoric stress components show inversion at fiber orientation angles near the *magic angle* in the FRE composites. These results are useful in soft robotic applications and in the biomechanics of fiber reinforced tissues such as the myocardium, arteries and skin.

## 1. Introduction

Soft bodied animals, such as sea anemones, octopi, and caterpillars, have an extraordinary ability to maintain posture, squeeze through narrow spaces, and move using forces transmitted by muscles to a hydrostatic skeleton [1]. Differentially arranged muscles undergo changes in tension-length relationships that are communicated to the animal skeleton to produce movement (**Figure 1a, b**). Additional geometric factors, for example those related to the helical organization of collagen fibers, control the diameter and length of the cylindrical skeleton in nematode worms to achieve the desired deformations during muscle actuation [2]. A fiber winding angle of 54.7° in the worm bodies, referred to as the *magic angle*, results in a maximum enclosed cylindrical volume [3]. Fiber reinforcement is also a characteristic in the mechanical behaviors of tissues like arteries and the myocardium that undergo anisotropic and large deformations at low stretches but exhibit strain stiffening at higher extensions [4,5].

Soft robotic applications demand materials that are soft and flexible to facilitate interactions with humans, have differential material properties to accommodate anisotropic load requirements, and undergo large stretch. McKibben actuators, comprised of a cylindrical expanding tube with woven fabric mesh, are designed to produce force-length relationships that are representative of biological muscles. Stroke actuation is generated *via* changes to the tube length upon pressurization [6]. Unlike the hydrostatic skeletons of animals, these actuators are designed to work in the linear strain regime and do not undergo dynamic variations in material properties during loading.

Bioinspired, tailored, fiber reinforced elastomers (FRE) exploit the ability to change their overall shape and material stiffness under loads; these are both potentially useful attributes in soft robotic applications. Development of such materials warrant a better understanding of the nonlinear force-displacement relationships in FRE materials and their links to fiber orientations. Methods using linear elasticity are unfortunately inadequate to describe the nonlinear, anisotropic, and large deformations in fiber reinforced elastomeric materials.

A macroscopic description of nonlinear material responses in FRE materials is given in terms of a strain energy density function per unit volume, $\Psi$, and described using strain invariants in the context of hyperelasticity. The number of invariants in the constitutive equation depend on existing material symmetries present in the FRE due to the underlying microstructure. For example, two invariants are required to describe an isotropic material, five for a transversely isotropic material, and nine for materials with orthotropic symmetry. Theoretical treatments in the development of constitutive formulations for fiber reinforced materials have explored invariant formulations, involving a subclass of invariants, and homogenization methods to investigate the development of instabilities [7,8]. Microstructure evolution during loading also influences the overall material response caused by a loss in the strong ellipticity condition due to the presence of macroscopic instabilities [9]. Multiaxial experimental data, required to assess the goodness of these models and test the underlying assumptions, are unfortunately scarce for FRE materials with controlled hierarchical microstructures.

This work uses a theoretical framework in combination with experimental methods to quantify the mechanics of FRE materials in transverse isotropic layouts. There are three specific goals of this study: First, to fabricate FRE materials with varying fiber angles in transversely isotropic layouts. Second, to use uniaxial and equibiaxial tests to assess the form of the constitutive model in FRE materials. Finally, we show an inversion point in the deviatoric stress components at the *magic angle*. Fiber winding at the *magic angle* is also associated with a minimization of the shear strains. The methods established in this study are essential in the design of dielectric elastomers with differential stiffness, electrically actuated materials as muscle actuators in robotics, and in sensing applications [10-12]. Such methods may also be a useful first step in characterizing the biomechanics of materials like the myocardium and skin that have fibers in transversely isotropic arrangements.

## 2. Materials and Methods

### 2.1. Fabrication of FRE composites

The thermally curable silicone, poly dimethyl siloxane (PDMS, Sylgard® 184, Dow Corning), was fabricated by mixing vinyl-terminated PDMS chains (part A) with a mixture of methyl hydrosiloxane copolymer chains with a platinum catalyst and an inhibitor (part B). The individual viscous constituents were mixed in a 30 (part A):1 (part B) ratio and degassed to remove bubbles [13]. We implemented a method involving a filament winding technique to orient commercially available polyester thread, mounted on an L frame, on a rectangular acrylic mandrel spinning at 1.7 Hz using a helically geared motor (**Supplementary Video S1**). The L-frame was translated using a linear actuator at a speed of 10mm/s to control the spacing between individual fibers during the winding process. Tension in the thread during winding was controlled using a spring-loaded ring. The PDMS mixture was poured on each of the rectangular faces of the mandrel following filament winding and the faces were cured in an oven at 60 ºC for 8 hours. Four large sheets of FRE composites, obtained using this method, were cut at various angles and tested using uniaxial (ASTM D412-06A) and equibiaxial experiments.

### 2.2. Material characterization

The fiber orientations, lengths, and pitch between adjacent fibers in FRE specimens were quantified using NIH ImageJ [14] and MATLAB 2016a (MathWorks, Inc, Natick, MA, USA) (**Table 1, Figure 1f**). Changes to the fiber orientations in the FRE composite during mechanical tests were quantified using Hough transforms.

### 2.3. Mechanical testing of FRE composites

We used a planar biaxial stretching instrument (BiSS Pvt. Ltd, Bangalore, India) for all uniaxial and equibiaxial mechanical experiments. Specific details of the stretcher and system working are given elsewhere [15]. The instrument consists of four independently controlled translating arms arranged along two perpendicular axes. Load cells (Model WMC-5, 2270 gm, 8.6 mN resolution, Interface Inc., Scottsdale, AZ, USA), located at the end of the arm in each axis, measured the forces ($F_1$, $F_2$) during stretching. The specimen thickness (t) was measured using callipers by gently sandwiching the specimen between two glass slides and the sample length ($L_1$) and width ($L_2$) were quantified using image processing. A video extensometer unit, consisting of a video camera (Sony HDR XR500) placed directly above the specimen, was used to quantify the in-plane stretches ($\lambda_1, \lambda_2$) in the specimens by tracking the displacements of four markers located in a central region of the sample. Particles were tracked and the in-plane Green strains ($E_{11}$, $E_{22}$, $E_{12}$) were calculated using the interpolation method from components of the deformation gradient tensor. The Cauchy stresses ($\sigma_{11}, \sigma_{22}$) were calculated as

$$\begin{aligned} \sigma_{11} &= \lambda_1 \frac{F_1}{tL_2} \\ \sigma_{22} &= \lambda_2 \frac{F_2}{tL_1} \end{aligned} \qquad (1)$$

Cauchy stress-Green strain data for the FRE composites and control silicone specimens were obtained using uniaxial and equibiaxial displacement-controlled protocols at displacement rate of 0.5 mm/s.

**2.4. Fitting Constitutive Model to Experimental Data**

Experimental stress-strain data were used to fit the form of strain energy function, Ψ, using a constrained nonlinear optimization algorithm implemented in MATLAB using the *fmincon* function. Differences between experimental Cauchy stresses in each direction ($\sigma_{ii}^{exp}; i = 1,2$) and the theoretically calculated stresses ($\sigma_{ii}^{calc}; i = 1,2$) are given by

$$Error = \sum_{j=1}^{N}(\sigma_{11j}^{calc} - \sigma_{11j}^{exp})^2 + (\sigma_{22j}^{calc} - \sigma_{22j}^{exp})^2 \qquad (2)$$

The index j refers to each point in the loading cycle. The error term was minimized with respect to material parameters using the specific form SEF for FRE composites. About ten different randomly generated initial guesses were used to establish that the converged solutions correspond to the global minimum. The reported solutions from the optimization method correspond when the tolerance placed on the objective function was less than 1E-6.[16]

## 3. Results and Discussion

### 3.1. Mechanics of FRE composites assessed using uniaxial experiments

Composite FRE samples were fabricated using a filament winding technique with polyester fibers in transversely isotropic layouts within a silicone elastomer (polydimethylsiloxane; PDMS; Sylgard®184, Dow Corning, USA) matrix mixed in the 30:1 ratio (**Figure 1c**). Composite FRE were cut at fiber orientations varying from 0° to 90° into dumbbell shaped test specimens using the ASTM D412-06A standard and tested uniaxially to failure (**Figure 1d**). Sheets of cured silicone material were also cut to the same dimensions and used as controls in the study. The filament winding technique used for the sample fabrication significantly reduced the dispersion in fiber angles and helped create a uniform microstructure in the specimen (**Supplementary Video S1**). In addition to the uniaxial experiments, square sheets of control silicone, and 15° and 30° oriented FRE samples (60 mm x 60 mm) were cut and subjected to equibiaxial displacement controlled protocols (**Figure 1e**). Microscopy images of the FRE specimens show that fibers have hierarchical substructures that aid with their binding to the matrix (**Figure 1f**).

Samples were clamped to a custom planar biaxial instrument and tested to failure. Stress-strain plots clearly show the effects of fiber orientations on the uniaxial mechanical properties of FRE samples (**Figure 2a**). Samples with fiber orientations aligned in the direction of the stretch (0°) were stiffer as compared to 90° oriented specimens. Data from a single polyester thread included in the figure, shows remarkable overlap with the results from the 0° oriented specimen. Both these samples showed little extensibility although the loads before failure were very high. The higher stiffness, corresponding to contributions from fibers to the overall material response, increased during stretch with fiber alignment in the other samples. The response of 90° FRE samples, comprised of fibers oriented perpendicular to the loading direction, was similar to that of the matrix silicone material. We calculated Young's moduli of the custom fabricated samples using experimental data ranging in the linear region upto 5% of the stress–strain curve (**Table 1**). The control silicone samples had moduli of 140.56 ± 13.24 kPa. [13] The moduli of FRE composites were bounded between the values for the silicone matrix and the polyester fibers. Specimens with fiber orientations near the direction of loading (~ 0°,15°) had significantly higher moduli due to the dominant contributions of the fibers. In contrast, samples with fibers oriented perpendicular to the loading direction had moduli values that were similar to the matrix.

The strain energy function, $\psi$, for a homogeneous, fiber reinforced, nonlinearly elastic, solid composite with transversely isotropic layout is given in terms of five strain invariants. [17] Assuming material incompressibility, we write the dependence of $\Psi$ on a subclass of invariants, $I_1$ and $I_4$, for the FRE sample by neglecting the invariants $I_2$ and $I_5$ for mathematical simplicity (**Supporting Information - Theoretical considerations**). $I_1$ is the trace of the Cauchy-Green strain tensor, $\boldsymbol{C} = \boldsymbol{F}^T \boldsymbol{F}$, and $I_4$ represents the directional dependence due to the fiber reinforcement, described as the square of the fiber stretch, included in $\Psi$. $\boldsymbol{F}$ is the deformation gradient and $^T$ denotes transpose. The in-plane components of $\boldsymbol{F}$ include stretch and shear terms to capture the effects of fiber reinforcement.

$$I_1 = tr(\boldsymbol{C}); \quad I_4 = \boldsymbol{M} \cdot \boldsymbol{CM} = \boldsymbol{m}^2 \tag{3}$$

$\boldsymbol{M} = (cos\,\theta, sin\,\theta, 0)^T$ is a unit vector defining the fiber orientation in the FRE. The angle $\theta$ represents fiber orientation along the loading direction (**Figure 1c**).

We used experimental stress-strain data from the uniaxial testing of FRE composites to quantify the variations in $I_1$ and $I_4$ with applied stretch, $\lambda$. **Figure 2b** shows a monotonic increase in $I_1$ with $\lambda$ for all samples oriented from 0-90° with respect to the loading direction. Sensitivity of $I_1$ to stretch is distinctly visible for the 15º, 30º and 45º FRE specimens; these responses are however indiscernible for the other fiber-reinforced specimens (**Figure 2b)**. In contrast, variations in $I_4$ show a decreased contribution as fibers become perpendicular to the loading direction up to angles of 60º relative to the stretch direction (**Figure 2c**). The values of $I_4$ are less than 1 and decreased with increased stretch for the 75º and 90º oriented samples. These data clearly show that the matrix plays a dominant contribution in the mechanics of large deformation in FRE composites. The experimental variations in $I_1$ and $I_4$ for specimens with varying fiber orientations in this study provide useful insights regarding the specific roles of fibers and their orientations in reinforced nonlinear solids that have largely been discussed in the literature based on analytical approaches.

We used these experimental data to assess the applicability of the inextensibility criterion that is frequently used as an additional constraint in $\Psi$ for the FRE composite specimens. Specifically, we test if line elements along the fiber undergo rigid rotations alone using the experimental data on FRE composites in our study. This inextensibility condition, first proposed by Adkins and Rivlin, is expressed as [18]

$$\left(\frac{ds}{ds\prime}\right)^2 = \lambda_1^2 cos^2\theta + \lambda_2^2 sin^2\theta = 1 \qquad (4)$$

where $ds$ and $ds\prime$ are the lengths of the line elements in the reference and deformed configurations respectively. $\theta$ is the angle with respect to the loading direction of fibers in the FRE samples (**Figure 1d**). $\lambda_i; i = 1:2$ are stretches in the principal directions. The kinematical constraint in Equation 4 is written in terms of Green strains in the fiber direction as

$$E_{fiber} = E_{11}cos^2\theta + 2E_{12}sin\theta cos\theta + E_{22}sin^2\theta = 0 \qquad (5)$$

We plot variations in $E_{fiber}$ with respect to the applied stretch for uniaxial testing specimens with different fiber orientations (**Figure 2d**). These data show that the Green strains in the fiber direction ($E_{fiber}$) are non-zero and follow a similar trend as that for $I_4$ (**Figure 2c**). Based on these data we conclude that the fibers are extensible and contribute significantly to the mechanical properties of the FRE composites. That is, inclusion of an additional constraint to enforce inextensibility in the form of $\psi$ is unsuitable for FRE material in our study.

We used Hough transform to characterize the dynamic changes in the fiber angles within the gage region of the FRE composites during uniaxial stretching experiments [19]. This method works by uniquely mapping each of the lines in the image, represented as a point (x,y), to a sine curve in parametric Hough space using coordinates (r,ϕ). The point to curve transition is described by $r = x \cos \varphi + y \sin \varphi$ where $\varphi \in [0, \pi)$ is the angle of the normal and r is the algebraic distance from the origin. **Figure 3a** shows that changes to the fiber angles in FRE composites in this study were relatively sensitive to the direction of loading. 0º, 15 º and 90º samples showed negligible changes in fiber angles to increasing load. The 90º sample underwent large stretch before specimen failure. These results are in good agreement with those of fiber extensibility of FRE specimens in **Figure 2d**. Because the fibers are extensible, they undergo both rigid rotations and elongation under stretch.

Failure of the composites occurred at a significantly high stress as compared to the control silicone samples and was associated with debonding of the fibers at the fiber-matrix interface. Fibers have hierarchical microstructures and include sub-fibrils that are critical to the binding of the fiber to the matrix (**Figure 1f**). Fiber debonding from the elastomer matrix resulted in local micro-failures that were characterized by local fluctuations in the fiber orientations with increased stretch. For example, the 75º FRE composite shows well bonded fibers to the matrix at the start of the experiment (**Figure 3b: location 1**). The **location 2** in **Figure 3b** clearly shows micro-failures and fiber pull-out from the matrix. Additional stretching of the samples resulted in several fiber pull-outs that are highlighted using arrows in **location 3 (Figure 3b) (Supplementary Video S2)**.

**3.2. Novel constitutive model for FRE composites**

We computed gradients of the strain energy function, Ψ(**C**), as a function of invariants $\left(\psi_i = \frac{\partial \Psi}{\partial I_i} ; i = 1, 4\right)$ using Cauchy stress- Green strain data from uniaxial experiments for the 45° oriented FRE test sample (**Supporting Information - Theoretical Considerations**). **Figure 4** shows variations in $\psi_1$ and $\psi_4$ as a function of $I_1$ and $I_4$. These observations show that $\psi_1$ monotonically increases with $I_1$ and $I_4$ in contrast to $\psi_4$ variations that are nonlinear and more sensitive to the degree of stretch. Fiber contributions decreased as they became perpendicular to the loading direction at higher extensions. These plots demonstrate the importance of coupling in the $I_1$ and $I_4$ terms that is crucial to the form of $\Psi$. Based on these results, we propose the following functional form for $\Psi$:

$$\Psi = c_1(I_1 - 3) + c_2(I_1 - 3)^2 + c_3(I_1 - 3)^2(I_4 - 1) + c_4(I_4 - 1)^2 + c_5(I_4 - 1)^3 \qquad (6)$$

where $c_j$; j=1:5 are constants that may be determined using experimental results. Equation 6 shows the individual contributions of the matrix, described in terms of $I_1$, contributions from the fiber stretch, $I_4$, and coupled terms in $I_1$ and $I_4$ that describe the matrix-fiber interactions in the nonlinear mechanics of reinforced FRE materials.

Data from the uniaxial testing experiments, with fiber orientations varying from 0° to 90°, were used to fit to the proposed form of $\Psi$ using the form which was obtained for the 45° specimen (**Figure 5a**). The unknown constants to the model (Equation 6) were obtained using a constrained optimization algorithm (*fmincon*) implemented in MATLAB. About ten different randomly generated initial guesses were used to obtain the converged solutions for each specimen. The optimization was terminated when the tolerance placed on the objective function was less than 1E-6. $r^2$ values greater than 0.9 are reported in this work (**Table 2).** Restrictions on $c_1$ & $c_2 \geq 0$ and $\Psi_1$ & $\Psi_4$>0 ensure polyconvexity of $\Psi$. The values of constants in Equation 6 ($c_j$; j=1:5) are different for specimens with the different fiber orientations with respect to the loading direction. **Table 2a** shows that values of constants $c_4$ & $c_5$ are higher for specimens with fiber orientations aligned in the direction of the stretch (0°, 15°) as compared to those oriented away from the loading direction.

We next used stress-strain results from equibiaxial experiments (15° and 30° specimens) to explore the predictive capability of our model (**Figure 5b**). FRE composites failed at relatively lower stretch ($\lambda_{1,2} \epsilon [1, 1.25]$) under equi-biaxial stretch as compared to the uniaxially tested samples **[Supplementary Video S3]**. Data from equibiaxial experiments demonstrate stiffening as fibers get aligned with any of the principal axes of stretch in the FRE composites. The constants to $\Psi$ for equibiaxial stretch conditions are clearly different as compared to the uniaxial data due to the different boundary conditions in equibiaxial tests; albeit the form of $\Psi$ in Equation 6 fits the uniaxial and equibiaxial results equally well (**Table 2b)**.

The methods adopted in this study for the formulation of the form of the strain energy function were pioneered by Rivlin and Saunders in studies on the large elastic deformation of rubber-like materials.[20] These were subsequently adapted by others to study the constitutive properties of fiber reinforced biological materials like the myocardium and arteries.[21,22,23] Existing forms of strain energy functions to describe the reinforcing contribution of fibers include an additive decomposition based on a neo-Hookean form for the contribution of the matrix, and a second term to describe the contribution of the fibers [24,25,26,27]. A sub-class of invariants, based on $I_1$ and $I_4$, is hence a natural choice for $\Psi$ in materials with transverse isotropic symmetry. Qui and Pence used analytical methods to show that a standard reinforcing model, with $I_4$ dependence included as a quadratic term, best describes the tensile contributions of fibers to the material response of fiber reinforced materials [28]. Humphrey and co-workers used a cubic dependence of $I_4$ and linear coupling between $I_1$ and $I_4$ to characterize the constitutive properties of the myocardium described using a transversely isotropic symmetry [22,29]. The Gasser-Holzapfel-Ogden model uses quadratic $I_4$ dependence in the exponential term for the phenomenological form of function used to describe arterial properties [30]. Experimental results from fibers in designed orientations within elastomeric matrices in our study show that coupled terms in $I_1$ and $I_4$ are crucial in determining the overall mechanical response of FRE composites.

## 3.3. Role of fiber orientation in mechanical design of bio-inspired materials

Angular orientations of fibers are important determinants in the mechanical response and overall shape of fiber reinforced materials seen abundantly in nature, such as in nematode bodies and octopi, that are commonly called muscular hydrostats [31,32]. Clark and Cowey used purely geometric considerations to show that the maximum enclosed volume of the cylinder, consisting of a helical reinforcing fiber of fixed length, was obtained at a pitch angle of $\theta_m$=54.7°.[33] Collagen fibers in the adventitial layer of arteries are wound around the axial axis in an almost symmetrical pattern with pitch angles ~53° and −51° that results in a quasi-isotropic mechanical response in arteries [34].

We computed variations in shear stress components in the FRE specimens for various fiber orientations using experimental stress-strain results from uniaxial experiments. Fiber orientations at the *magic angle* reduce the possibility of failure by debonding in the FRE composites [35]. To compute the effects of shear stress on the fiber orientation effects, we computed shear stress components in the fiber direction as a function of the applied fiber stretch given by [36]

$$3\,\boldsymbol{M}.\boldsymbol{\sigma}\,\boldsymbol{M} - (tr(\boldsymbol{\sigma})) = c \qquad (7)$$

where c is a constant that depends on the applied fiber stretch and $\boldsymbol{M} = (\cos\theta,\ \sin\theta,\ 0)^T$ is a unit vector along the fiber direction. We used constants obtained for the 45° FRE composite in Equation 6 to evaluate the components of the deviatoric stress components along various possible fiber orientation angles, $0° \leq \theta \leq 90°$, for applied stretches, $1.1 \leq \lambda \leq 1.5$, under uniaxial tensile loads. **Figure 6a** shows that the deviatoric stresses change nonlinearly with stretch but have linear dependence at values near the magic angle, $\theta_m$. The variations in shear stress with fiber orientations may be delineated into three distinct domains given by

$$3\,\boldsymbol{M}.\boldsymbol{\sigma}\,\boldsymbol{M} - (tr(\boldsymbol{\sigma})) = \begin{cases} > 0\ for\ 0 < \theta < \theta_m \\ = 0\ for\ \theta = \theta_m \\ < 0\ for\ \theta_m < \theta < \frac{\pi}{2} \end{cases} \qquad (8)$$

Recent theoretical investigations suggest that the *magic angle* acts as an inversion point corresponding to the point of extremum strain in the material under loading [36,37]. To test this hypothesis, we computed the deviatoric stress components (Equation 7) using material parameters obtained using optimization algorithms for FRE specimens with various different fiber orientations (**Table 2a**) and plotted the results in **Figure 6b**. These data show that most samples, barring the 0° specimen, have an inversion point, $\theta_a$, in the range of 54.6°-62.5° (**Table 2a**). These observations provide a clear connection between fiber winding angles on the mechanical response and the constitutive relations for FRE materials.

## 4. Conclusions

Studies on fiber reinforced elastomers have primarily explored analytical formulations to describe their stress-strain properties [9,38]. Experimental results to test the veracity and efficacy in implementing such models are however scarce. We fabricated bio-inspired fiber reinforced FRE materials, composed of polyester fibers embedded in a silicone matrix, and characterized the nonlinear mechanics of FRE composites using a theoretically motivated continuum mechanical framework. The filament winding technique permits uniform fiber arrangement in the specimens. Uniaxial and equibiaxial results from FRE samples, oriented at different angles with the load direction, show the individual contributions of fibers and the matrix to the form of the strain energy function. We used these data to propose a new form of the strain energy function that demonstrates the importance of including a coupling term in invariants $I_1$ and $I_4$ to better describe the stress-strain properties of the FRE composites. We use these data to show an inversion in the deviatoric components of stress at fiber orientations at ~54.7° which has been described as the *magic angle*.

For simplicity, we have ignored the effects of damage such as debonding between the fibers and the matrix over the range of deformations in our study. The new constitutive model (Equation 4) can be implemented in any numerical scheme to study the effects of complex loading and boundary value conditions for flexible FRE composite materials. The data in this study are unique because they provide a robust method to test many of the analytical models for FRE materials that have several potential applications in biomechanics and in soft robotics. These studies are a first step to investigating the mechanics of active elastomers, and in electrically actuated fibers, such as Nitinol, in matrices that may be used to mimic the muscular hydrostatic skeletons in engineering applications.

## 5. Acknowledgements

NG gratefully acknowledges financial support from STC-ISRO (ISTC241) and intramural support from the Indian Institute of Science, Bangalore for funding this project. We also thank Mr. B. Kiran Kumar who helped in the fabrication of the fiber reinforced elastomers and performed the mechanical experiments reported in this study.

## 6. Conflict of Interest

None of the listed authors have any potential conflict with the results reported in this work.

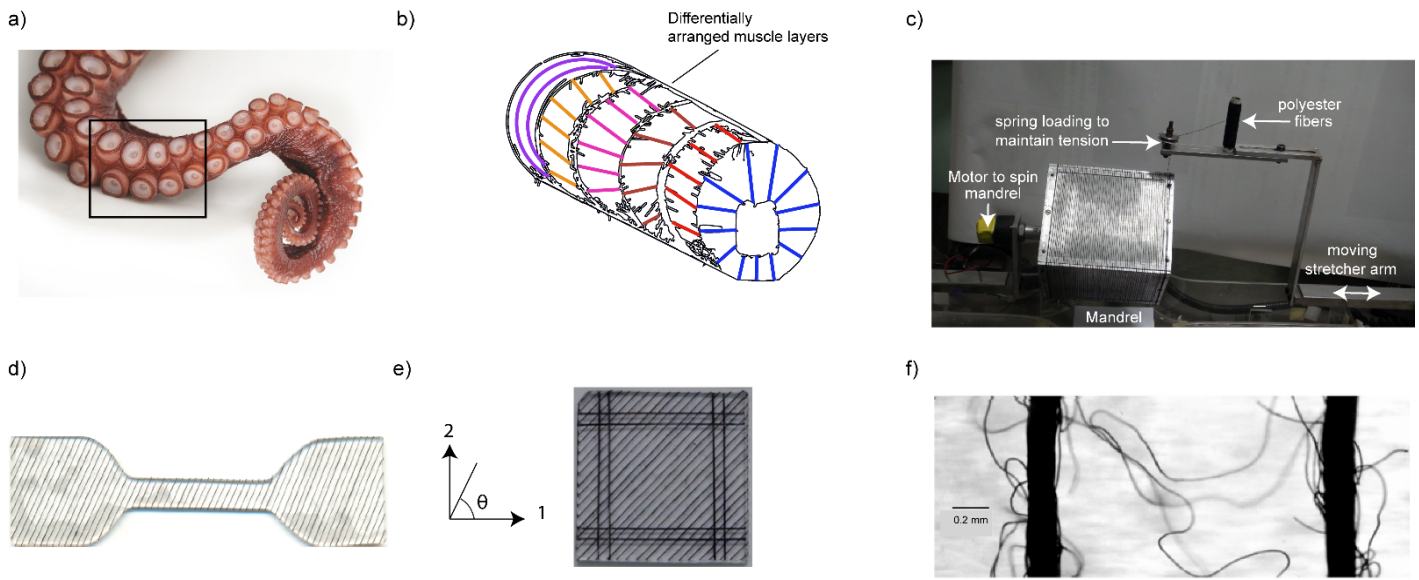

**Figure 1.** a) A coiled and twisted octopus arm as an example of muscular hydrostat. [shutterstock.com]  b) Cartoon shows differentially arranged muscle layers in a region of the octopus arm. Fiber orientations change through the arm length and help control movements. c) Schematic shows the custom developed system used to fabricate the novel FRE specimens in our study. d & e) FRE specimens were cut at different fiber orientations for the uniaxial and equibiaxial tests. A representative co-ordinate system used in the measurement of the fiber angles with loading direction is shown. f) The polyester fibers in FRE composites have hierarchical microstructures that aid in the adhesion of the fibers to the underlying matrix in FRE composites.

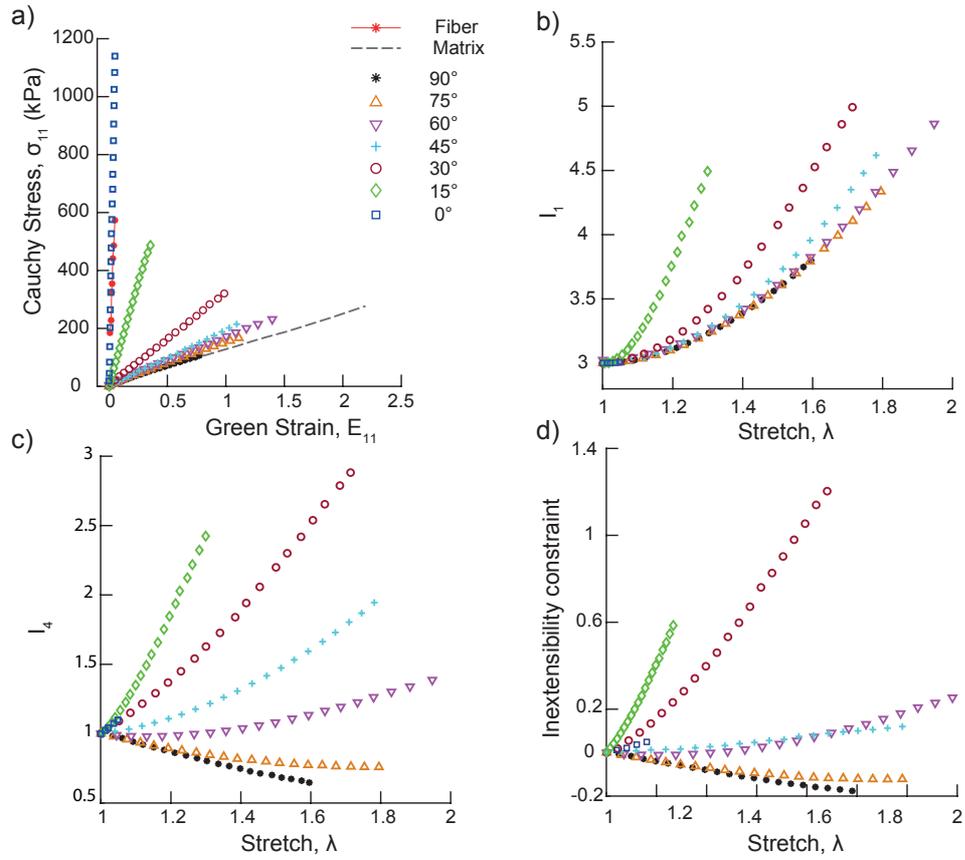

**Figure 2.** FRE samples with fibers at various orientations with respect to the loading direction were tested uniaxially to failure. **a)** Cauchy stress - Green strain data are plotted for the FRE samples, control silicone matrix sample, and a polyester (fiber) thread for FRE specimens oriented at varying angles with the loading direction. **b)** and **c)** Variations in invariants $I_1$ and $I_4$ with increased uniaxial stretch are shown for the FRE specimens under uniaxial loading. **d)** Variations in the inextensibility constraint, $E_{fiber}$, in Equation 5 with experimentally measured stretches are shown under different initial orientations. The components of $E_{fiber}$ are non-zero with stretch which show that fibers extensibility is important in the mechanical behaviors of the FRE composites.

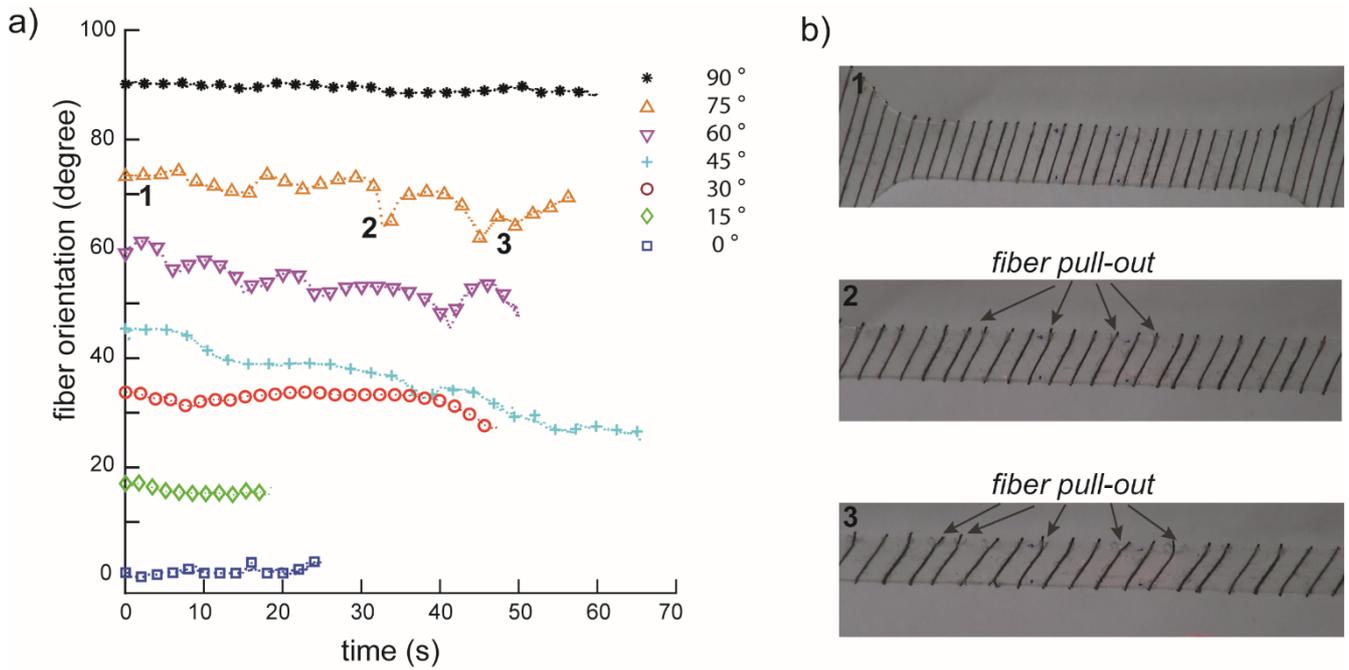

**Figure 3.** Dynamic changes in the fiber orientations occur under uniaxial loading of FRE samples. **a)** Changes in the fiber orientation angles are plotted for each of the different FRE specimens with loading time. **b)** Fiber debonding from the elastomer matrix correlated with local fluctuations in fiber orientational changes in a representative sample corresponding to 75º FRE. **Location 1** shows well bonded fibers to the matrix at the beginning of the experiment. **Location 2** shows presence of micro-failures and fiber pull-out from the matrix. Failures increased with higher stretch in the sample (**location 3**).

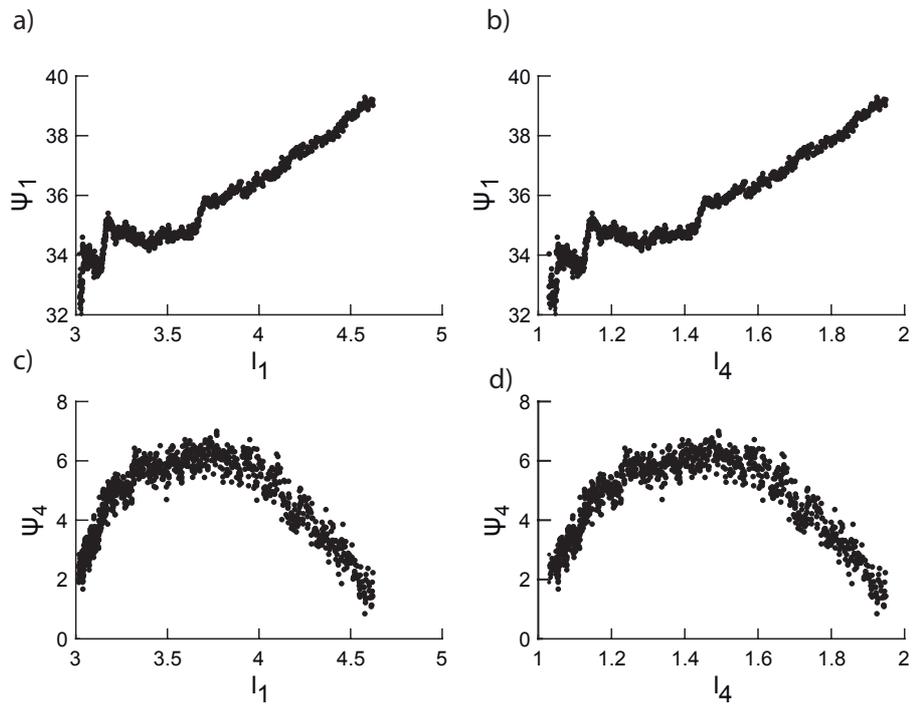

**Figure 4.** Results from stress-strain experiments for the 45° FRE sample were used to compute the variations in gradients of the strain energy function with invariants, $I_1$ and $I_4$. **a)** and **b)** show that $\psi_1$ increases monotonically with $I_1$ and $I_4$. In contrast, **c)** and **d)** show that variations in $\psi_4$ are nonlinear and more sensitive to the degree of stretch in the sample.

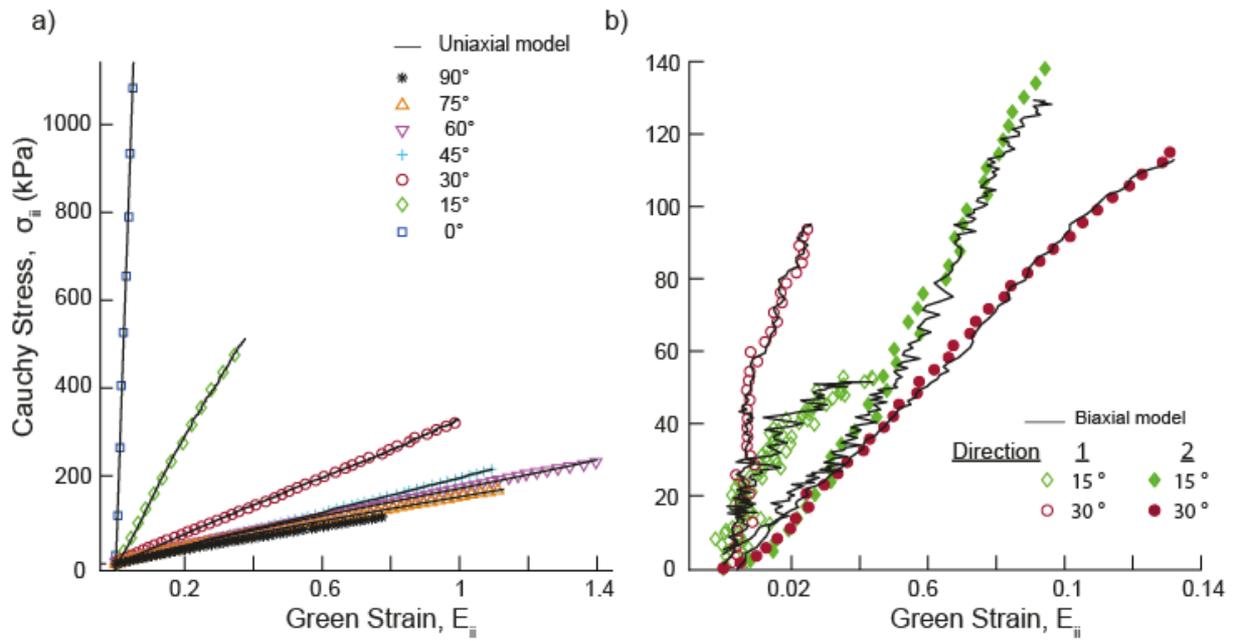

**Figure 5.** Experimental stress-strain data from FRE samples were used to obtain the unknown coefficients in the strain energy function (Equation 6). **a)** The model fit the uniaxial test results for FRE samples tested at different orientations with $r^2>0.9$. **b)** equibiaxial tensile tests for 15° and 30° oriented FRE composites with $r^2>0.98$ (Table 2).

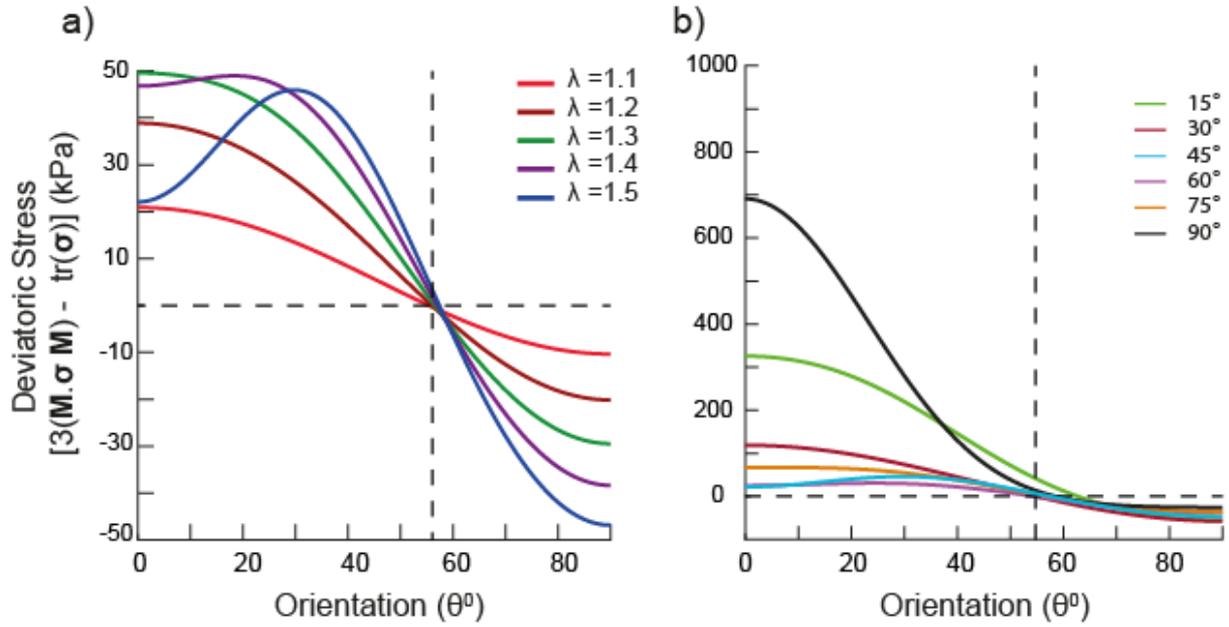

**Figure 6. a)** Variations in the deviatoric stress components with fiber orientation were calculated using model fits to the 45° FRE specimen under uniaxial stretch for different cases of stretch, λ. The magnitude of deviatoric stress components are a minimum at $\theta_m$. **b)** Variations in the deviatoric stresses were calculated using experimental results for all FRE samples varying and are shown for the different orientations. FRE specimens, barring the 0° sample, show an inversion point $54° \leq \theta_a \leq 63°$ (Table 2).

**Table 1.** Characterization of the FRE composites. Specimens were imaged and the fiber angles and distance between individual fibers were quantified. Uniaxial stress-strain data were used to calculate the specimen moduli for all FRE samples in the study.

| | Specimen | Fiber Angle (°) (Mean ± Std) | Spacing Between Fibers (mm) (Mean ± Std) | Elastic Modulus (kPa) |
|---|---|---|---|---|
| UNIAXIAL | 0° | 1.25 ± 0.05 | 1.69 ± 0.03 | 22573.1 |
| | 15° | 13.41 ± 0.73 | 1.54 ± 0.13 | 1621.5 |
| | 30° | 29.73 ± 0.69 | 1.67 ± 0.09 | 357.1 |
| | 45° | 43.3 ± 0.86 | 1.71 ± 0.16 | 229.6 |
| | 60° | 58.8 ± 0.49 | 1.75 ± 0.12 | 211.6 |
| | 75° | 75.31 ± 0.66 | 1.78 ± 0.21 | 202.5 |
| | 90° | 90.08 ± 0.6 | 1.78 ± 0.21 | 172.6 |

**Table 2.** Unknown coefficients to the strain energy function (Equation 6) were obtained by fitting computed stresses from the constitutive model to experimental data from **a)** uniaxial and **b)** equibiaxial experimental data. The inversion point, $\theta_a$, corresponding to a minimum in the deviatoric stress value (Equation 7) are included for uniaxially tested samples.

a) Uniaxial FRE Specimens

|  | $c_1$ | $c_2$ | $c_3$ | $c_4$ | $c_5$ | $r^2$ value | $\theta_a °$ |
|---|---|---|---|---|---|---|---|
| 0° | 0.2 | 590.5 | -27.3 | 2816.9 | -2008.9 | 0.90 | - |
| 15° | 0.05 | 0.04 | 0.01 | 83.26 | -21.24 | 0.98 | 62.5 |
| 30° | 53.91 | 0 | -0.03 | 0.84 | -0.06 | 0.96 | 54.9 |
| 45° | 34.31 | 0 | 1.71 | 14.16 | -11.63 | 0.99 | 57.4 |
| 60° | 35.36 | 0 | 0.29 | 0.49 | -3.88 | 0.99 | 54.6 |
| 75° | 31.37 | 1.32 | 6.26 | 6.37 | -5.36 | 0.99 | 57.3 |
| 90° | 26.61 | 0.38 | 0.24 | 14.07 | 37.51 | 0.98 | 58.5 |

b) Biaxial FRE Specimens

|  | $c_1$ | $c_2$ | $c_3$ | $c_4$ | $c_5$ | $r^2$ value |
|---|---|---|---|---|---|---|
| 15° | 94 | 3358 | -19701 | 59 | -42 | 0.98 |
| 30° | 100.3 | 325.8 | -4580.7 | 123.6 | 213.5 | 0.99 |

**Supplementary Video 1: Custom set-up for the fabrication of FRE Specimens using a filament winding technique.**

**Supplementary Video 2: Gage region of a uniaxial tested 75°oriented sample is shown. Fiber debonding from the matrix plays an important role in the failure of the FRE composites.**

**Supplementary Video 3: Equibiaxial tensile testing of 30° oriented FRE composite to failure.**

**8. Supplementary Information**

**8.1 Theoretical Considerations**

We use a linear decomposition of the strain energy function (SEF) into an isotropic component, based on the matrix material, and an anisotropic component related to the transversely isotropic fiber layout in the matrix. [1,2]

*8.1.1 Kinematics of Pure Homogeneous Deformation of a Thin Sheet:* Let $\varphi: \Omega_0 \to R^3$ describe the mapping of a material point X from a reference configuration $\Omega_0$ of a body to its deformed configuration $\Omega$ by $x = \chi(X, t)$. Upper case letters denote material description whereas lower case the spatial description; both are described using Cartesian coordinates. The deformation gradient is given by $F = \partial x / \partial X$, where J=det(**F**) is the Jacobian of the transformation.

The generalised description of the deformation gradient for a specimen subjected to biaxial deformations in the (x,y) plane (**Figure 1e**) are given by

$$[F] = \begin{bmatrix} F_{11} & F_{12} & 0 \\ F_{21} & F_{22} & 0 \\ 0 & 0 & F_{33} \end{bmatrix} \quad (S1)$$

$F_{11}$, $F_{22}$ refer to the in-plane components based on stretches $\lambda_1$ and $\lambda_2$. $F_{33} = \lambda_3$ is the stretch orthogonal to the plane. Off-diagonal components, $F_{12}$ and $F_{21}$, in the expression represent the shear components that are essential when considering the effects of fibers. We define $C = F^T F$ and $b = F F^T$ as the right and left Cauchy-Green strain tensors where [T] denotes transpose.

We assume that all fibers in the FRE composites are oriented in single direction in reference configuration at angle $\theta$ with respect to the direction 1 (**Figure 1d**). Fiber orientation vectors in the reference (**M**) and deformed (**m**) configurations are given by

$$\mathbf{M} = (\cos\theta, \sin\theta, 0)^T \quad \& \quad \mathbf{m} = (F_{11}\cos\theta + F_{12}\sin\theta, F_{21}\cos\theta + F_{22}\sin\theta, 0)^T \quad (S2)$$

The strain energy function, $\psi$, for the FRE composites with a transversely isotropic layout is given in terms of invariants of the Cauchy Green strain tensor, **C**, by

$$\psi = \hat{\psi}[I_1(\mathbf{C}), I_2(\mathbf{C}), I_3(\mathbf{C}), I_4(\mathbf{C}), I_5(\mathbf{C})] \quad (S3)$$

The three principal invariants are given by

$$I_1 = tr(\mathbf{C}), I_2 = \frac{1}{2}[(tr\mathbf{C})^2 - tr(\mathbf{C}^2)], I_3 = det(\mathbf{C}) \quad (S4)$$

where, *tr* and *det* refer to the trace and determinant operations respectively. We define additional pseudo-invariants, $I_4$ and $I_5$, based on the fiber direction, **M**.

$$I_4 = \mathbf{M} \cdot \mathbf{CM} = \mathbf{m}^2 \quad \& \quad I_5 = \mathbf{M} \cdot \mathbf{C}^2\mathbf{M} = \mathbf{m}.\,\mathbf{bm} \quad (S5)$$

$I_4$ is the square of stretch in the fiber direction and is a function of fiber length. In contrast, $I_5$ depends on the changes in the fiber length and shear strains. Strain invariants for pure homogeneous extension of a thin and incompressible FRE sheet are given in terms of the stretches, $\lambda_i$ ($i = 1,2,3$), as

$$I_1 = \lambda_1^2 + \lambda_2^2 + \lambda_3^2; \; I_2 = \lambda_1^2\lambda_2^2 + \lambda_2^2\lambda_3^2 + \lambda_1^2\lambda_3^2; \; I_3 = (\lambda_1\lambda_2\lambda_3)^2 = 1 \quad (S6)$$

We have used $I_3 = 1$, that is $\lambda_3 = \frac{1}{\lambda_1\lambda_2}$, based on the incompressibility assumption. In the absence of residual stresses, the Cauchy stresses for the incompressible FRE composites are given as

$$\boldsymbol{\sigma} = -p\mathbf{I} + 2J^{-1}\mathbf{F}\frac{\partial\hat{\psi}(\mathbf{C})}{\partial\mathbf{C}}\mathbf{F}^T \quad (S7)$$

$p$, is a hydrostatic term present in the expression due to the incompressibility assumption and is determined using boundary conditions and the equilibrium equations.

In terms of the strain invariants, we write

$$\boldsymbol{\sigma} = -p\boldsymbol{I} + 2\hat{\psi}_1\boldsymbol{b} + 2\hat{\psi}_2(I_1\boldsymbol{b} - \boldsymbol{b}^2) + 2\hat{\psi}_4\boldsymbol{m}\otimes\boldsymbol{m} + 2\hat{\psi}_5(\boldsymbol{m}\otimes\mathbf{bm} + \mathbf{bm}\otimes\boldsymbol{m}) \quad \text{(S8)}$$

where $\hat{\psi}_i = \partial\hat{\psi}/\partial I_i$, $i = 1,2,4,5$.

*8.1.2 Determination of Strain Energy Function:* We use the method pioneered by Rivlin and Saunders [3] for large elastic deformation of rubber like materials and adapted by others for biological materials [4,5,6] to obtain the form of $\widehat{\Psi}(\boldsymbol{C})$. We neglect effect of fiber extensibility and consider a subclass of invariants that depend on invariants $I_1$ and $I_4$ alone in the form of $\widetilde{\Psi}(\boldsymbol{C})$ for mathematical simplicity. Cauchy stresses are hence given by

$$\boldsymbol{\sigma} = -p\boldsymbol{I} + 2\left\{\frac{\partial\widetilde{\Psi}}{\partial I_1}\boldsymbol{b} + \frac{\partial\widetilde{\Psi}}{\partial I_4}\boldsymbol{m}\otimes\boldsymbol{m}\right\} \quad \text{(S9)}$$

In component form, the Cauchy are written using kinematic equations as

$$\sigma_{11} = -p + 2\left[\widetilde{\Psi}_1\lambda_1^2 + \widetilde{\Psi}_4(\lambda_1^2\cos^2\theta)\right] \quad \text{(S10)}$$

$$\sigma_{22} = -p + 2\left[\widetilde{\Psi}_1\lambda_2^2 + \widetilde{\Psi}_4(\lambda_2^2\sin^2\theta)\right]$$

$$\sigma_{33} = -p + 2\widetilde{\Psi}_1(\lambda_1^{-2}\lambda_2^{-2})$$

where $\widetilde{\Psi}_i$ denotes partial derivative of $\widetilde{\Psi}$ with respect to invariant, $I_i$. The Lagrange multiplier, p, may be eliminated using plane stress condition $\sigma_{33} = 0$ for thin incompressible FRE sheet.

$$\sigma_{11} = 2\left\{\widetilde{\Psi}_1\left(\lambda_1^2 - \frac{1}{\lambda_1^2\lambda_2^2}\right) + \widetilde{\Psi}_4\lambda_1^2\cos^2\theta\right\} \quad \text{(S11)}$$

$$\sigma_{22} = 2\left\{\widetilde{\Psi}_1\left(\lambda_2^2 - \frac{1}{\lambda_1^2\lambda_2^2}\right) + \widetilde{\Psi}_4\lambda_2^2\sin^2\theta\right\}$$

We solve these two equations for variables $\widetilde{\Psi}_1$ and $\widetilde{\Psi}_4$.

$$\widetilde{\Psi}_1 = \frac{\sigma_{11}\lambda_2^2\sin^2\theta - \sigma_{22}\lambda_1^2\cos^2\theta}{\lambda_2^2\sin^2\theta\left(\lambda_1^2 - \frac{1}{\lambda_1^2\lambda_2^2}\right) - \lambda_1^2\cos^2\theta\left(\lambda_2^2 - \frac{1}{\lambda_1^2\lambda_2^2}\right)} \quad \text{(S12)}$$

$$\widetilde{\Psi}_4 = \frac{\sigma_{22}\left(\lambda_1^2 - \frac{1}{\lambda_1^2\lambda_2^2}\right) - \sigma_{11}\left(\lambda_2^2 - \frac{1}{\lambda_1^2\lambda_2^2}\right)}{\lambda_2^2\sin^2\theta\left(\lambda_1^2 - \frac{1}{\lambda_1^2\lambda_2^2}\right) - \lambda_1^2\cos^2\theta\left(\lambda_2^2 - \frac{1}{\lambda_1^2\lambda_2^2}\right)}$$

*8.1.3 Restrictions on the Unknown Material Coefficients:* We use the form of the strain energy function obtained from the 45° FRE specimen to fit the uniaxial stress-strain data for FRE specimens with different fiber orientations based on the kinematic equations, $\lambda_1 = \lambda$; $\lambda_2 = \lambda_3 = 1/\sqrt{\lambda}$. We impose restrictions to the possible values of the unknown material coefficients in the expression for $\widetilde{\Psi}$ based on the physically possible material response. The coefficients corresponding to terms $(I_1 - 3)$ and $(I_1 - 3)^2$ are positive because they represent contributions from the matrix and have equivalent description of shear modulus. To enforce polyconvexity of the strain energy function, we include additional restrictions: $\Psi_1$ & $\Psi_4 > 0$.

We test that the term $\tau_{11} dE_{11} + d\tau_{22} dE_{22} > 0$ is satisfied for all the FRE samples that were tested in our study. $d\tau$ is the change in the stress corresponding to any infinitesimal logarithmic strain change $dE$. This expression corresponds to the Drucker stability criterion which guarantees positive definiteness of the strain energy density function for planar deformation. The coefficients reported in the study satisfied this restriction. Such a formulation of the strain energy function may be easily implemented in any finite element framework to study complex geometries and boundary value problems.